\documentclass[11pt]{article}
\pdfoutput=1

\usepackage{jheppubmod}

\usepackage[utf8]{inputenc}
\usepackage{amsmath, amsfonts, amstext, amssymb, mathrsfs, mathtools}
\usepackage{graphicx}
\usepackage{comment}
\usepackage{dutchcal}
\usepackage{feynmp-auto}
\usepackage{color}

\newcommand{\secref}[1]{Sec. \ref{#1}}
\newcommand{\appref}[1]{Appendix \ref{#1}}
\newcommand{\figref}[1]{Fig. \ref{#1}}
\newcommand\mat[4]{{\left( \begin{array}{cc} {#1} & {#2} \\ {#3} & {#4} \end{array} \right)}}

\begin{document}

\title{Quantum gravity on the black hole horizon}
\author{Nava Gaddam, }
\author{Nico Groenenboom, and}
\emailAdd{gaddam@uu.nl}
\emailAdd{n.groenenboom@uu.nl}
\author{Gerard 't Hooft}
\emailAdd{g.thooft@uu.nl}
\affiliation{Institute for Theoretical Physics and Center for Extreme Matter and Emergent Phenomena, Utrecht University, 3508 TD Utrecht, The Netherlands.}
\date{\today}

\abstract{We study scattering on the black hole horizon in a partial wave basis, with an impact parameter of the order of the Schwarzschild radius or less. This resembles the strong gravity regime where quantum gravitational effects appear. The scattering is governed by an infinite number of virtual gravitons exchanged on the horizon. Remarkably, they can all be summed non-perturbatively in $\hbar$ and $\gamma \sim M_{Pl}/M_{BH}$. These results generalise those obtained from studying gravitational backreaction. Unlike in the eikonal calculations in flat space, the relevant centre of mass energy of the collisions is not necessarily Planckian; instead it is easily satisfied, $s \gg \gamma^2 M^2_{Pl}$, for semi-classical black holes. The calculation lends further support to the scattering matrix approach to quantum black holes, and is a second-quantised generalisation of the same.}

\maketitle

\section{Introduction}
Inelastic collision of high energy bodies is expected to result in production of black holes. Moreover, at large impact parameters (compared to the Schwarzschild radius), the collapse is expected to be described by classical physics, dominated by eikonalised single graviton exchange \cite{Banks:1999gd, Eardley:2002re}. Of course, Hawking famously claimed that semi-classical physics results in information loss in the final state \cite{Hawking:1976ra}. This may be seen as a thermalisation of the initial states among the large number of degrees of freedom associated to the intermediate macroscopic black hole. Smaller impact parameters and presumably complicated quantum corrections are often accused of hiding the mysteries of information loss in all aspects of scattering after collapse, in the evaporation process \cite{Giddings:2007qq}.

Interestingly enough, the cross-section of the inelastic collisions producing black holes resembles scattering of the initial states on an existing, already formed Schwarzschild background \cite{Amati:1987wq}. This is owed to the fact that, during the collapse of a spherical shell for instance, the (apparent) horizon forms long before the shell has fallen past the Schwarzschild radius ($R_S$) of the eventual black hole. Therefore, one expects to capture a part of the physics associated with scattering in flat space with impact parameter of the order of the Schwarzschild radius by studying scattering on the intermediate black hole state (already at a perturbative level). The conventional eikonal approximation \cite{Levy:1969cr} has been used to study physics (at impact parameters larger than $R_S$) in flat space gravitational scattering \cite{tHooft:1987vrq, Amati:1987uf, Kabat:1992tb}. In that case, the scattering energies are necessarily Planckian $\sqrt{s} \gg M_{Pl}$ and the impact parameter larger than the Schwarzschild radius $b\gg R_S$. 

In the present article, we consider scattering near the Schwarzschild horizon in a partial wave basis. Therefore, the impact parameter is necessarily of the order of the Schwarzschild radius or less: $b\leq R_S$. We work in an approximation where transverse momentum transfer on the sphere is small, implying $b\gg L_{Pl}$. There are three scales left in the game: the centre of mass energy of the scattering (say $\sqrt{s}$), the mass of the intermediate black hole state $M_{BH}$, and the Planck mass $M_{Pl}$. From these, two dimensionless parameters arise, say $\gamma \coloneqq \sqrt{8\pi G_N} / R_S \sim M_{Pl}/M_{BH}$ and $s/(\gamma^2 M^2_{Pl})$. Remarkably we find a result that is non-perturbative in $\gamma$, by summing over an infinite number of graviton-exchanging ladder diagrams. This implies that the result is also non-perturbatively resummed in $\hbar$ because the ladder diagrams include infinitely many loop contributions. At every order in $\gamma$ in the ladder diagrams, we work to leading order in $s \gg \gamma^2 M^2_{Pl}$. However, unlike the conventional eikonal approximation, this is far easier to satisfy for black holes much bigger than Planck size. For a solar mass black hole, for instance, the center of mass energy required for this approximation is $\sqrt{s}\gg 10^{-38} M_{Pl}$; this is already satisfied for all the massive particles in the standard model with their rest mass alone. Moreover, the energy of particles diverges near the horizon, validating this approximation further. Therefore, this is indeed the strong gravity regime where the mysteries of black hole information loss must necessarily be resolved.

We compute a four-point correlation function of matter fields on the Schwarzschild horizon. This includes multi-particle states owing to the fact that we work with fields in a partial wave basis. The correlator captures scattering processes in two different channels (much like in familiar nucleon-nucleon scattering), both exchanging virtual gravitons. The result is surprisingly simple and lends support to the S-matrix approach for quantum black holes and generalises earlier results in \cite{Hooft:2015jea, Hooft:2016itl, Betzios:2016yaq, Betzios:2020wcv}. Furthermore, this work naturally provides a completely second-quantised description of the earlier quantum mechanical results. Since the black hole geometry is locally weakly curved, one may not expect this calculation to be very instructive. However, as the scattering matrix maps asymptotic states, very rich physics emerges that is characteristically different to that of flat space. For instance, we find that the amplitude is entirely dominated by low angular momentum modes; in contrast, the flat space amplitude is dominated by very large angular momentum modes. While certain numerical factors require detailed calculations which we leave to a companion article \cite{Gaddam:2020mwe}, the conceptual idea and the results (up to explicit numerical factors) can be obtained from symmetry principles as we show in this paper. The theory under consideration is in \secref{sec:theory}, the ladder diagrams to be computed in \secref{sec:diagrams}, and a discussion on the results in \secref{sec:discussion}.

\section{The action}\label{sec:theory}
We consider the Einstein-Hilbert action coupled minimally to matter, say a massless scalar field for simplicity:
\begin{equation}
    S ~ = ~ \dfrac{1}{2} \int \sqrt{-g} \, \mathrm{d}^4x \left(\dfrac{R}{8 \pi G_N} - ~ \nabla_\mu \phi \nabla^\mu \phi\right) \, . \nonumber
\end{equation}
We focus on the fluctuations about a spherically symmetric background $\left(g_{\mu\nu} \rightarrow g^0_{\mu\nu} + \kappa h_{\mu\nu}\right)$ and a vanishing scalar profile, where
\begin{equation}
    g^0_{\mu\nu} ~ = ~ - 2 A\left(x,y\right) \mathrm{d}x\mathrm{d}y + r\left(x,y\right) \mathrm{d}\Omega^2_2 \, , \nonumber
\end{equation}
where $\mathrm{d}\Omega^2_2$ is the round metric on the two-sphere, and $\kappa = \sqrt{8 \pi G_N}$. For the Schwarzschild background, we have that $A = \exp\left(1 - r/R_S\right)R_S/r$ and $r$ is defined via 
\begin{equation}
xy ~ = ~ 2 R^2_S \left(1 - r/R_S\right)\exp\left(r/R_S - 1\right) 
\end{equation}
in the familiar Kruskal-Szekeres coordinates. While we will later focus on the vacuum Schwarzschild solution in this note, we present a method that applies to generic spherically symmetric solutions supported by matter. Of course, the fluctuations $h_{\mu\nu}$ are not all physical and the field needs to be gauge-fixed, an issue we now turn to.

\subsection{Gauge fixing}
The gauge that best exploits the spherical symmetry of the problem at hand is that of Regge and Wheeler \cite{Regge:1957td}. The metric fluctuations are first expanded in spherical harmonics and split into odd and even parity modes $h_{\mu\nu} = \sum_{l, m} h^+_{lm, \mu\nu} + \sum_{l, m} h^-_{lm, \mu\nu}$ where a symmetric two dimensional field $H_{ab}$ with indices $a,b \in \{x,y\}$ and a scalar field $K$ are introduced. Both $H_{ab}$ and $K$ are only functions of $x,y$ (and of course depend on the suppressed $l,m$ indices) since the angular dependence is extracted out in the spherical harmonic expansion; we label the angular coordinates by capitalised indices $A,B$. In the Regge-Wheeler gauge, the even mode is then given by
\begin{align}
    h^+_{lm, \mu\nu} ~ = ~ \mat{H_{ab} Y_{lm}}{0}{0}{r^2 g_{AB} K Y_{lm}} \, , \nonumber
\end{align}
and the explicit form of the odd mode will turn out to be irrelevant (as we will describe below). Of the six physical degrees of freedom in the metric fluctuations, four sit in the even mode above, and two in the odd-parity mode $h^-_{lm, \mu\nu}$. Conveniently, the choice of gauge results in a decoupling of the odd and even parity modes in the quadratic action. Moreover, spherical symmetry of the background ensures a decoupling between the different partial waves at quadratic order. 

Finally, a comment on Faddeev-Popov ghosts is in order. We will assume that they contribute to the scattering to be considered only at sub-leading order in the $s\gg \gamma^2 M^2_{Pl}$ limit; this is familiar from \cite{Kabat:1992tb}. A further comment on this matter is due in \secref{sec:nearhorizon}.\\

\subsection{An effective two-dimensional theory}\label{sec:effective2dtheory}
The spherical symmetry of the background and the decoupling between the different parity modes allows for the theory to be reduced on the sphere to arrive at an effective two-dimensional one \cite{Martel:2005ir}. Since the two-dimensional metric is conformally flat, a Weyl rescaling of the form $g_{ab} \rightarrow A\left(x,y\right) \eta_{ab}$ allows for a trade of the curvature for additional potential terms. Furthermore, the field re-definitions $\mathfrak{h}_{ab} = r A\left(x,y\right) H_{ab}$, $\mathcal{K} = r K$ allow us to finally write the resulting effective two-dimensional action where the space-time indices are raised and lowered by the flat metric $\eta_{ab}$ owing to the Weyl rescaling:
\begin{align}\label{eqn:2daction}
    S_{2d} ~ &= ~ \sum_{l,m}\dfrac{1}{4}\int \mathrm{d}^2x \left(\mathfrak{h}^{ab}_{lm} \Delta^{-1}_{abcd}\mathfrak{h}^{cd}_{lm} + \mathfrak{h}^{ab}_{lm} \Delta^{-1}_{L,ab} \mathcal{K}_{lm} + \mathcal{K}_{lm} \Delta^{-1}_{R,ab} \mathfrak{h}^{ab}_{lm} + \mathcal{K}_{lm} \Delta^{-1} \mathcal{K}_{lm}  \right. \nonumber \\
    &\qquad\qquad \left. + 2 \phi_{lm} \left(\partial^2 - \dfrac{\partial^2 r}{r}\right) \phi_{lm} - 2 \phi_{lm} \left(\dfrac{A \ell \left(\ell + 1\right)}{r^2}\right) \phi_{lm} + \gamma \, \mathfrak{h}^{ab}_{lm} \partial_a \phi_0 \partial_b \phi_{lm}\right) \, ,
\end{align}
The operators appearing in this action are explicitly defined in \appref{sec:quadopers}. The interaction vertex \textit{a priori} contains two terms with the even-parity graviton of the form $\mathcal{K}\eta^{AB} T_{AB}$ and $\mathfrak{h}^{ab} T_{ab}$, and a similar term containing the odd-parity graviton. However, we first note that the four point function of interest is always built out of two three-point vertices in this theory. We consider in-going particles in the collision to carry momentum along the coordinate $x$ while the out-going ones along $y$. The odd-parity graviton only couples to transverse matter momenta. That transverse momentum (along the angular coordinates) effects are sub-leading implies that the interaction term contains no odd-parity graviton modes \cite{Verlinde:1991iu, Gaddam:2020mwe}. Furthermore, the assumptions that $s \gg \gamma^2 M^2_{Pl}$ and $b\gg L_{Pl}$ have two important consequences: the first is that all diagrams containing the field $\mathcal{K}$, and those with momentum exchange are all sub-leading at every order in $\gamma$, while the second is that the off-diagonal components of $\mathfrak{h}^{ab}$ do not contribute. Moreover, in the scattering process $\phi_{\ell_1 m_1}\phi_{\ell_2 m_2}\rightarrow\phi_{\ell_3 m_3}\phi_{\ell_4 m_4}$ being considered, spherical symmetry of the background implies that the angular momentum is either transferred across the vertex (the \textit{transfer} channel) or conserved along the vertex (the \textit{conserved} channel) but not distributed across the fields with Clebsch-Gordon coefficients; so, we label one of the fields in the interaction vertex as $\phi_0$ and require it to be with some fixed angular momentum. Those diagrams that distribute angular momentum will introduce interactions between the various partial waves and are consequently sub-leading \cite{Verlinde:1991iu}. These facts have the consequence that the interaction vertex simplifies to be of the form displayed in \eqref{eqn:2daction}. It is worth noting that this form ignores virtual graviton exchanges with $\mathfrak{h}^{ab}_{\ell=0}$ in the transfer channel unless the external particles are all in the $\ell=0$ state. Including these interactions will also introduce interactions between the various partial waves and are therefore expected to be sub-leading.

The propagators arising from the above action are defined by
\begin{equation}\label{eqn:propagatordefs}
    \begin{pmatrix}
        \Delta^{-1}_{abcd} & \Delta^{-1}_{L,ab} \\ 
        \Delta^{-1}_{R,cd} & \Delta^{-1}
    \end{pmatrix}
    \begin{pmatrix}
        \mathcal{P}^{cdef} & \mathcal{P}^{cd}_R \\ 
        \mathcal{P}^{ef}_L & \mathcal{P}
    \end{pmatrix}
    =  
    \begin{pmatrix}
        \delta^{ef}_{ab} & 0 \\ 
        0 & 1
    \end{pmatrix}
    \delta^2\left(x-x^\prime\right)
    \, , \nonumber
\end{equation}
with $\Delta^{-1}_{abcd}\Delta^{cdef} = \delta^{ef}_{ab} \delta^{(2)}(x-x^\prime)$ and $\Delta^{-1}\Delta = \delta^{(2)}(x-x^\prime)$. These operators are difficult to invert in full generality. In what follows, we will discuss the special case of the two-dimensional surface being the horizon of the Schwarzschild black hole.

\subsubsection{Physics near the black hole horizon}\label{sec:nearhorizon}
\begin{figure}[htb]
	\includegraphics[width=0.45\textwidth]{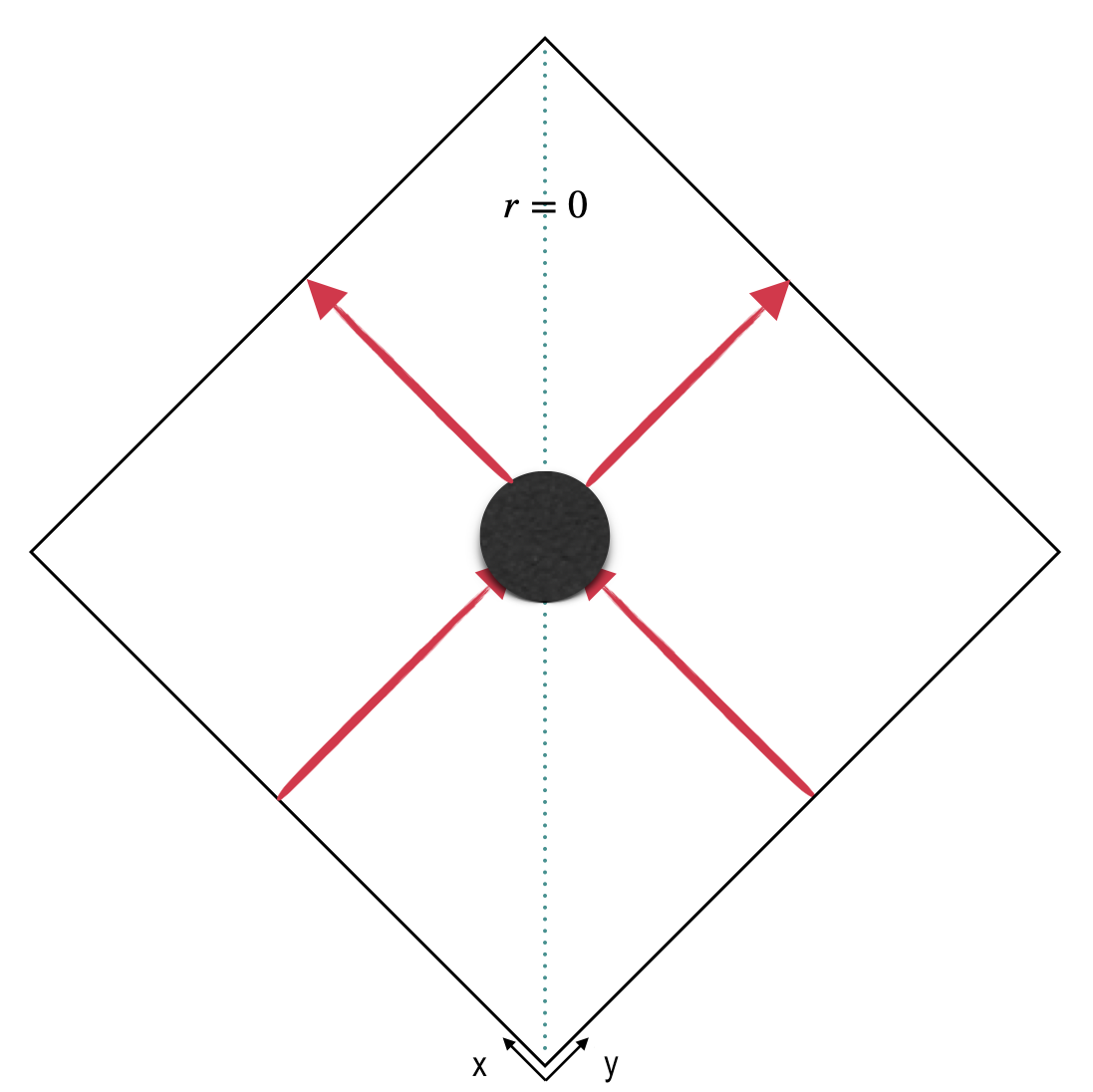}
	\centering
	\caption{Two particles fall in to the black hole from the asymptotic past and two particles emerge and escape to future infinity. This scattering amplitude is mediated by graviton fluctuations of the horizon, which are ignored in Hawking's calculation. Because the interactions take place near the horizon, the impact parameter is at best of the order of $R_{S}$} \label{fig:2to2}
\end{figure}

As discussed in the Introduction, scattering processes in a black hole background naturally capture quantum gravity effects perturbatively in $\gamma$. Hawking's free-field theory result shows that the modes of quantum fields on past and future null infinities are related by Bogoliubov coefficients which have thermal pre-factors. Gravitational interactions, governed by $\gamma$, alter this picture. In this article, we will be able to obtain results for $2\rightarrow 2$ scattering non-perturbatively in $\gamma$. In \figref{fig:2to2}, we show a schematic representation of the $2\rightarrow 2$ scattering of interest. When two particles fall into a black hole from past infinity, we will compute the amplitude of two particles escaping to future infinity; this amplitude is mediated by the metric fluctuations of the horizon.

When thought of in Eddington-Finkelstein coordinates, the asymptotic future of all in-going wave fronts from past null infinity $\mathscr{I}^-$ coincides with the asymptotic past of the wave fronts emanating from future null infinity $\mathscr{I}^+$. In a collapsing problem, near this region, an apparent horizon forms well before the collapsing shell cross the Schwarzschild radius. Therefore, strong quantum gravitational physics which is expected in scattering with impact parameter $b \leq R_S$ is evidently captured by this region. 

Alternatively, in Kruskal-Szekeres coordinates, this may be seen as the region where the lightcone coordinates $x,y < R_S$. In this approximation, we see that the $A(x,y) \sim 1 + \mathcal{O}(xy)$ and $r(x,y) \sim R_S + \mathcal{O}(xy)$, to linear order. As one moves ever closer to the diamond at Planckian distances, precise methods may be possible but this is beyond the present paper. We focus on $L_{Pl} \ll b \leq R_S$.

The fact that the two-dimensional theory is conformally flat implies that the tensorial propagator is of the form
\begin{equation}
    \mathcal{P}^{abcd} ~ = ~ \dfrac{1}{4} f_\ell\left(k^2\right) \left(\eta^{ac}\eta^{bd} + \eta^{ad}\eta^{bc}\right) \, .
\end{equation}
The factor of a quarter is merely chosen for convenience, and is arbitrary. A very lengthy calculation shows that the above form factor for the Schwarzschild space-time is given by \cite{Gaddam:2020mwe}
\begin{equation}\label{eqn:formfactor}
    f_{\ell}\left(k^2\right) =  - \dfrac{4 R^2_S}{\left(\lambda+1\right)} - \dfrac{2 R^4_S k^4}{\left(\lambda + 1\right)\left(\lambda - 3\right)\left(k^2 + R^{-2}_S \lambda\right)} \, ,
\end{equation}
where we defined the constant $\lambda = \ell^2 + \ell + 1$. The explicit form of this function $f_{\ell}(k^{2})$ is largely unimportant for the present paper; it will turn out that only $f_{\ell}(0)$ will turn out to play a role. Neverttheless, this propagator deserves attention. First, we observe that it displays the familiar ultraviolet divergences in the $k \rightarrow \infty$ limit. In the soft-limit, which is of interest because the $s\gg \gamma^2 M^2_{Pl}$ limit is dominated by diagrams with no momentum exchange, it is finite; the horizon ensures infrared regularity. This may be interpreted as an effective two-dimensional mass for the graviton mode $\mathfrak{h}^{ab}$ on the horizon; it is worth noting that the four-dimensional graviton remains massless and this is merely a curvature artefact on the horizon. Although irrelevant for the rest of this article, the momentum dependent term is rather curious. The $\ell = 0$ mode causes a change in sign in the term. This mode is very interesting as it corresponds to change in black hole mass \cite{Zerilli:1970se}. Worse still, the $\ell = 1$ mode has a pole; indeed, equations of motion cannot be inverted for this mode and it can be reduced to terms containing only the odd-parity graviton \cite{Zerilli:1970se, Nagar:2005ea} and might be considered to affect the centre of mass motion of the black hole. Both of these modes suggest additional gauge redundancy. Although we have fixed the Regge-Wheeler gauge, resulting ghosts have not been accounted for, as mentioned earlier. At sub-leading order in the $s\gg \gamma^2 M^2_{Pl}$ limit, where momentum exchange diagrams will be of importance, these additional gauge redundancies must be addressed. 

Bringing together all the notes discussed so far, we will now focus on a given partial wave $\ell , m$ and suppress these indices. The action \eqref{eqn:2daction} near the horizon now reduces to
\begin{align}\label{eqn:horizonaction}
    S_{\text{hor}} ~ &= ~ \dfrac{1}{4} \int \mathrm{d}^2 k \,  \mathfrak{h}^{ab} \, \mathcal{P}^{-1}_{abcd} \, \mathfrak{h}^{cd} ~ - ~ \dfrac{1}{2} \int \mathrm{d}^2 p \, \phi \left(p^2 + \dfrac{\lambda}{R^2}\right) \phi \nonumber \\
    &\quad ~ + \gamma \int \mathrm{d}^2k \, \mathrm{d}^2p_1 \, \mathrm{d}^2p_2 \, \delta^{(2)}\left(k + p_1 + p_2\right) \mathfrak{h}_{ab}\left(k\right) p^a_1 p^b_2 \phi_0\left(p_1\right) \phi\left(p_2\right) 
\end{align}
for each partial wave, in momentum space. The absence of the $\mathcal{K}$ field is owed to the fact that it is a transverse field on the sphere, which gives a sub-dominant effect as expected from \cite{Verlinde:1991iu}. The Feynman rules for the contributing propagators are as follows:\\
\begin{fmffile}{feyn_propagators}
\begin{align}
  \begin{fmfgraph*}(90,0)
    \fmfleft{i1}
    \fmfright{i2}
    \fmf{phantom}{i1,i2}
    \fmfi{fermion,label=$\phi_{\ell m}(p)$,label.side=left}{vpath (__i1,__i2)}
  \end{fmfgraph*} ~~~ &= ~ \dfrac{-i}{p^2 + \frac{\lambda}{R^2_S} - i \epsilon} \nonumber \\
  \begin{fmfgraph*}(90,0)
    \fmfleft{i3}
    \fmfright{i4}
    \fmf{phantom}{i3,i4}
    \fmfi{dashes,label=$\phi_{0}(p)$,label.side=left}{vpath (__i3,__i4)}
  \end{fmfgraph*} ~~ &= ~ \dfrac{-i}{p^2 + \frac{1}{R^2_S} - i \epsilon} \nonumber \\
    \begin{fmfgraph*}(60,0)
    \fmfleft{i5}
    \fmfright{i6}
    \fmf{phantom}{i5,i6}
    \fmflabel{$\mathfrak{h}^{ab}_{\ell m}$}{i5}
    \fmflabel{$\mathfrak{h}^{cd}_{\ell m}$}{i6}
    \fmfi{curly,label=$\mathfrak{h}_{\ell m}(k)$,label.side=left}{vpath (__i5,__i6)}
  \end{fmfgraph*} ~ \qquad &= ~ 2 i \mathcal{P}_{abcd} \, , \nonumber \\ \nonumber
  \end{align}
\end{fmffile}
and the vertex is given by \\
\begin{fmffile}{feyn_vertex}
\begin{align}
  \begin{fmfgraph*}(100,60)
    \fmfleft{i8,i7}
    \fmfright{o2}
    \fmf{gluon}{i7,v1}
    \fmf{dashes_arrow}{i8,v1}
    \fmf{fermion}{v1,o2}
    \fmflabel{$\mathfrak{h}^{ab}_{\ell m}$}{i7}
    \fmflabel{$p_2$}{i8}
    \fmflabel{$p_1 ~ .$}{o2}
    \fmfdot{v1}
    \fmfv{label=$i \gamma ~ p^1_a ~ p^2_b$,label.angle=60,label.dist=20}{v1}
  \end{fmfgraph*} \nonumber
\end{align}
\end{fmffile}
Here, we included a symmetry factor of $2$ for the scalar propagators (owing to an exchange of external scalar legs) and a factor of $8$ for the graviton propagator to account for the $a\leftrightarrow b$ and $c\leftrightarrow d$ symmetry. The vertex drawn here contributes to the transfer channel, where the angular momentum is transferred across the virtual graviton. A similar vertex with two solid scalar lines contributes to the conserved channel.

\section{The non-perturbative amplitude}\label{sec:diagrams}
The tree-level transfer channel ladder diagram is\\
\begin{fmffile}{feyn_tree}
 \begin{center}
  \begin{fmfgraph*}(80,80)
    \fmfstraight
    \fmfleft{i1,i2}
    \fmfright{o1,o2}
    \fmflabel{$p_2$}{i1}
    \fmflabel{$p_2$}{o1}
    \fmflabel{$p_1$}{i2}
    \fmflabel{$p_1$}{o2}
    \fmf{dashes_arrow}{i1,v1}
    \fmf{fermion}{v1,o1}
    \fmf{fermion}{i2,v2}
    \fmf{dashes_arrow}{v2,o2}
    \fmf{gluon,label=$k=0$}{v1,v2}
  \end{fmfgraph*} 
 \end{center}
\end{fmffile}
The amplitude is straight forward to calculate and is given by $i\mathcal{M} = 4i s^2 \gamma^2 R^2_S / \left(\lambda + 1\right) = 4i s^2 \kappa^2 / \left(\lambda + 1\right)$, where we defined the Mandelstam variable that gives the center of mass energy of the collision: $s = -\frac{1}{2}\left(p_1 + p_2\right)^2 = p^1_x p^2_y$; this is owed to the fact that $p^1_y = 0 = p^2_x$ as discussed in \secref{sec:effective2dtheory}. Moving on to loop diagrams, a two loop diagram in the transfer channel is of the following form:
\begin{fmffile}{feyn_transfertwoloop}
 \begin{center}
  \begin{fmfgraph*}(200,90)
    \fmfstraight
    \fmfleft{i1,i2}
    \fmfright{o1,o2}
    \fmf{dashes_arrow}{i1,v1}
    \fmf{fermion}{v5,o1}
    \fmf{fermion}{i2,v2}
    \fmf{fermion}{v1,v3}
    \fmf{dashes}{v2,v4}
    \fmf{fermion}{v4,v6}
    \fmf{dashes}{v3,v5}
    \fmf{dashes_arrow}{v6,o2}
    \fmf{gluon, tension=0}{v1,v2}
    \fmf{gluon, tension=0}{v3,v4}
    \fmf{gluon, tension=0}{v5,v6}
  \end{fmfgraph*} \nonumber
 \end{center}
\end{fmffile}
There are six such diagrams with all possible attachments for the virtual gravitons. It is also evident that a one-loop diagram contributes to the conserved channel and not to the transfer channel. The solid line is not transferred across to the bottom; it stays on top instead. This is true of all odd-loop diagrams. Therefore, the transfer channel contributes only to the even loop diagrams and the conserved channel only to the odd ones. In a manner similar to that of \cite{Levy:1969cr}, the general order-$n$ amplitude can be written as
\begin{align}
    i\mathcal{M}_n &= \left(i s \gamma\right)^{2n} \int \prod_{j=1}^n \left(\frac{\mathrm{d}^2 k_j}{\left(2\pi\right)^2} i f_\ell\left(k_j\right)\right) \times I \times \left(2\pi\right)^2 \delta^{(2)}\left(\sum_{j=1}^n k_j\right) \, , \nonumber
\end{align}
where $I$ contains all possible matter propagators; these can be derived following the techniques of \cite{Levy:1969cr}. Noticeably, we have inserted a factor of $i s \gamma$ for each vertex following the assumption of \secref{sec:effective2dtheory} that internal vertices contributing to additional virtual graviton momenta are sub-leading at each order in perturbation theory. After some algebra \cite{Levy:1969cr, Gaddam:2020mwe}, the amplitude simplifies remarkably
\begin{align}\label{eqn:nlooptransferamplitude}
    i \mathcal{M}_n ~ = ~ i \dfrac{\gamma^2 s^2 \left(i\chi\right)^{n-1}}{n!} f_\ell\left(0\right) ~ = ~ 4s \dfrac{\left(i \chi\right)^n}{n!} \, ,
\end{align}
where we have defined $\chi = - \frac{1}{4} \gamma^2 s f_\ell\left(0\right)$. These expressions are, at a technical level, identical 2d analogs of those in \cite{Levy:1969cr} and hold for arbitrary $f_{\ell}(0)$. The nonperturbative transfer channel amplitude is now straightforward to write down:
\begin{equation}\label{eqn:transfernonpert}
    i \mathcal{M}_{\text{transfer}} ~ = ~ 4s \sum_{n ~ \text{odd}}^\infty \dfrac{\left(i \chi\right)^n}{n!} ~ = ~ 4 i s \sin\left(\chi\right) \, .
\end{equation}
A similar calculation returns an analogous result for the conserved channel: 
\begin{equation}
i \mathcal{M}_{\text{conserved}} ~ = ~ 4 s \left(\cos\left(\chi\right) - 1\right) \, .
\end{equation} 
The complete amplitude is therefore a sum of these two channels:
\begin{equation}\label{eqn:nonpertamplitude}
    i \mathcal{M} ~ = ~ 4s \left(e^{i \chi} - 1\right) \, .
\end{equation}
This result may also be seen as the four-point amplitude $\langle\phi_{\ell m}\left(p_2\right)\phi_0 \phi_{\ell m}\left(p_1\right)\phi_0\rangle$. Note that the only non-vanishing component of $p_1$ is along the $x$ direction, which is in-going into the black hole; similarly $p_2$ is only non-vanishing along $y$, which is out-going from the black hole. Therefore, the amplitude can be written as 
\begin{equation}
    \langle\phi_{\ell m}\left(p_{\text{out}}\right)\phi_0 \phi_{\ell m}\left(p_{\text{in}}\right)\phi_0\rangle \sim \left(e^{i \frac{\gamma^2 R^2_S}{\lambda + 1} p_{\text{in}} p_{\text{out}}} - 1\right) . \nonumber
\end{equation}
Fourier transforming the out-going field, we find that the amplitude is only non-vanishing when
\begin{equation}\label{eqn:in-out-relation}
    y_{\text{out}} ~ = ~ \tilde{\lambda} \, p_{\text{in}} \quad \text{with} \quad \tilde{\lambda} = \dfrac{8\pi G_N}{\ell^2 + \ell + 2} \, ,
\end{equation}
implying that $\langle\phi_{\ell m}\left(y_{\text{out}}\right)\phi_0 \phi_{\ell m}\left(p_{\text{in}}\right)\phi_0\rangle$ is given by
\begin{equation}
    \langle\phi_{\ell m}(\tilde{\lambda} p_{\text{in}})\phi_0 \phi_{\ell m}\left(p_{\text{in}}\right)\phi_0\rangle \, . \nonumber
\end{equation}
This result is remarkably similar to those obtained from a first quantised formalism in \cite{Hooft:2015jea, Hooft:2016itl, Betzios:2016yaq}, with an important difference in the pre-factor. At first sight, this may appear to be an inconsistency. However, it can be checked that both results are indeed correct. Moreover, a detailed analysis shows \cite{Gaddam:2020mwe} that restricting the graviton fluctuations to be of the form $\mathfrak{h}_{xx}\left(x,y\right) = \mathfrak{h}_{xx}\left(y\right)\delta\left(x\right)$ on the past horizon and $\mathfrak{h}_{yy}\left(x,y\right) = \mathfrak{h}_{yy}\left(x\right)\delta\left(y\right)$ on the future horizon results in the factor ($\ell^2 + \ell + 1$) found in the previous literature \cite{Hooft:2015jea, Hooft:2016itl, Betzios:2016yaq}, for the Dray-'t Hooft shockwave metric. The result of the present article is therefore a generalisation that includes arbitrary off-shell graviton fluctuations.

\section{Discussion}\label{sec:discussion}
In this article, we studied scattering processes with an impact parameter less than $R_S$ but greater than Planck scale. This is the regime where quantum gravity effects are important. Remarkably, the amplitude can be computed non-perturbatively in $\hbar$ and $\gamma \sim M_{Pl}/M_{BH}$, provided that the centre of mass energy of the collision satisfies $s\gg \gamma^2 M^2_{Pl}$; this relation is easily satisfied for large black holes. The non-perturbative amplitude captures many-particle scattering states as it is obtained for second quantised fields in a partial wave basis. It results in a remarkable relation between the in-going and out-going fields \eqref{eqn:in-out-relation} that is a generalisation of the analogous result in a first quantised formalism \cite{Hooft:2015jea, Hooft:2016itl, Betzios:2016yaq}. This lends further support to the scattering matrix approach to quantum black holes as a resolution to the information paradox for the external observer. Despite the expected large collision energies, information transfer is carried out by soft gravitons at all orders in perturbation theory.  This implies that no momentum is transferred to or from the infalling observer. In itself, the $2-2$ amplitude is expected to be suppressed. But the black hole eikonal phase we develop allows for $2-N$ amplitudes with particle production to be calculated; such high point functions are expected to resolve the information problem. 

\subsection{Shortcomings} 
We now list the limitations of this work. First, we have focussed on the Schwarzschild horizon as depicted in \figref{fig:2to2}. However, physics further away may be incorporated by studying the classical Regge-Wheeler potential \cite{BGP2020}. We have assumed that diagrams sub-leading in the $s \gg \gamma^2 M^2_{Pl}$ approximation can consistently be dropped at every loop order in $\gamma$; this has been proved to be consistent for quantum field theory in flat space \cite{Levy:1969cr}. While the scalar $\phi^3$ theory violates this consistency \cite{Eichten:1971kd}, interacting vector theories have been shown to respect it \cite{Tiktopoulos:1971hi}. A formal proof of consistency for the present case would be an interesting technical addition to the literature. We have assumed that contributions of ghosts are sub-leading in the $s \gg \gamma^2 M^2_{Pl}$ approximation. For eikonal graviton fluctuations around flat space, this has been shown to be true \cite{Kabat:1992tb}. In the case at hand, the non-perturbative resummation to arrive at \eqref{eqn:nonpertamplitude} enforces a soft limit $k=0$ for all exchanged virtual particles. This fact, combined with the discussion below the form factor \eqref{eqn:formfactor} lends support to this assumption. We have also neglected higher order terms in the graviton fluctuations. Some of these are presumably responsible for non-renormalisability of the theory in the UV and therefore expected to be sub-leading in the soft limit; we hope to analyse the others in the future. Finally, we have also assumed that all contributions arising from any interaction between the various partial waves are sub-leading; this is natural given the spherical symmetry of the Schwarzschild solution. An important consequence of this is that certain spherically symmetric graviton fluctuations $\mathfrak{h}^{ab}_{\ell=0}$ drop out of the calculations in some loop diagrams; incorporating these would shed light on the dynamical change in mass of the black hole.

\subsection{Future directions} 
The tools proposed in this article can be generalised to include various other standard model fields. Owing to the fact that the results are governed by the near horizon physics, they are agnostic of the asymptotic nature of the solution. Therefore, it may be of interest to understand the CFT calculation of the non-perturbative amplitude \eqref{eqn:nonpertamplitude} for small black holes in AdS/CFT. It is also interesting to repeat the analysis for charged and rotating black holes. The amplitude \eqref{eqn:nonpertamplitude} may provide an additional window into understanding the quantum chaotic nature of the spectrum of black hole microstates; it would be very interesting to adapt the boundary condition proposed in \cite{Betzios:2020wcv} to the present second quantised description. Including transverse momentum transfer may be tractable by including a $B$-field that is familiar from string theory \cite{tHooft:1989tkg, Verlinde:1991iu, Kiem:1995iy, deHaroOlle:1997hx, deHaro:1998tj}. Finally, implications for the infalling observer would be interesting to address.

\section*{Acknowledgements}
We are thankful to Panos Betzios, Anne Franzen, Steve Giddings, Olga Papadoulaki, and Tomislav Prokopec. This work is supported by the Delta-Institute for Theoretical Physics (D-ITP) that is funded by the Dutch Ministry of Education, Culture and Science (OCW).

\begin{appendix}

\section{Quadratic operators}\label{sec:quadopers}
The operators appearing in \eqref{eqn:2daction} are
\begin{align}
\Delta^{-1} ~ &= ~ -\partial^2+F_a^a \, , \\
\Delta^{-1}_{R,ab} ~ &= ~ - \eta_{ab} \left(\partial^2 + \dfrac{1}{2} \left(U^c - V^c\right) \partial_c + \dfrac{1}{2} \eta^{cd} \left(\mathcal{W}^A_{cd} - \mathcal{W}^r_{cd}\right) \right. \nonumber \\
&\qquad\qquad\qquad - \left.A\left(r\right) \dfrac{\ell(\ell+1)}{2r^2}\right) + \partial_a \partial_b + U_{(a} \partial_{b)} + \mathcal{W}^A_{ab} - F_{ab} \, ,  \\
\Delta^{-1}_{L,ab} ~&= ~ - \eta_{ab} \left(\partial^2 - \dfrac{1}{2} \left(U^c - V^c\right) \partial_c - A\left(r\right) \dfrac{\ell(\ell+1)}{2r^2} \right) + \partial_a \partial_b - U_{(a} \partial_{b)} - F_{ab} \, ,  \\
\Delta^{-1}_{abcd} ~ &= ~ \dfrac{1}{2} \eta_{ac} V_{[b} \partial_{d]} + \dfrac{1}{2} \eta_{bd} V_{[a} \partial_{c]} + \dfrac{1}{2} \eta_{ab} \left( V_{(c} \partial_{d)} + \mathcal{W}^r_{cd} + \dfrac{1}{2} V_c V_d\right) \nonumber \\
&\qquad + \dfrac{1}{2} \eta_{cd} \left(- V_{(a} \partial_{b)} + \dfrac{1}{2} V_a V_b\right) +  \eta_{ab} \eta_{cd} \left(\dfrac{1}{4} A\left(r\right) R_{2d} - \dfrac{1}{4} V^e U_e + A\left(r\right) \dfrac{\ell(\ell+1)}{2r^2}\right) \nonumber \\
&\qquad - \eta_{ac} \eta_{bd} \left(\dfrac{1}{2} A\left(r\right) R_{2d} - \dfrac{1}{2} V^e U_e + A\left(r\right) \dfrac{\ell(\ell+1)}{2r^2}\right) \, , 
\end{align}
where we defined the following fields
\begin{align}
V_a ~ &\coloneqq ~ 2 \partial_a \log r \, , \\
F_{ab} ~ &\coloneqq ~ \dfrac{1}{r} \tilde{\nabla}_a \tilde{\nabla}_b r ~ = ~ \dfrac{1}{2} \tilde{\nabla}_{(a} V_{b)} + \dfrac{1}{4} V_a V_b \, , \\
R_{2d} ~ &\coloneqq ~ - \dfrac{1}{A} \tilde{\Box} \log A \, , \\
\mathcal{W}^r_{ab} ~ &\coloneqq ~ \partial_{(a}V_{b)} \, , \\
\mathcal{W}^A_{ab} ~ &\coloneqq ~ \partial_{(a}U_{b)} \, .
\end{align}

\end{appendix}


\begin{thebibliography}{50}%
\makeatletter
    \bibitem{Banks:1999gd}
    T.~Banks and W.~Fischler,
    \textit{A Model for high-energy scattering in quantum gravity,}
    [arXiv:hep-th/9906038 [hep-th]].
    
    \bibitem{Eardley:2002re}
    D.~M.~Eardley and S.~B.~Giddings,
    \textit{Classical black hole production in high-energy collisions,}
    Phys. Rev. D \textbf{66} (2002), 044011
    doi:10.1103/PhysRevD.66.044011
    [arXiv:gr-qc/0201034 [gr-qc]].

    \bibitem{Hawking:1976ra}
    S.~W.~Hawking,
    \textit{Breakdown of Predictability in Gravitational Collapse,}
    Phys. Rev. D \textbf{14} (1976), 2460-2473
    doi:10.1103/PhysRevD.14.2460
    
    \bibitem{Giddings:2007qq}
    S.~B.~Giddings and M.~Srednicki,
    \textit{High-energy gravitational scattering and black hole resonances,}
    Phys. Rev. D \textbf{77} (2008), 085025
    doi:10.1103/PhysRevD.77.085025
    [arXiv:0711.5012 [hep-th]].
    
    \bibitem{Amati:1987wq}
    D.~Amati, M.~Ciafaloni and G.~Veneziano,
    \textit{Superstring Collisions at Planckian Energies,}
    Phys. Lett. B \textbf{197} (1987), 81
    doi:10.1016/0370-2693(87)90346-7
    
    \bibitem{Levy:1969cr}
    M.~Levy and J.~Sucher,
    \textit{Eikonal approximation in quantum field theory,}
    Phys. Rev. \textbf{186} (1969), 1656-1670
    doi:10.1103/PhysRev.186.1656
    
    \bibitem{tHooft:1987vrq}
    G.~'t Hooft,
    \textit{Graviton Dominance in Ultrahigh-Energy Scattering,}
    Phys. Lett. B \textbf{198} (1987), 61-63
    doi:10.1016/0370-2693(87)90159-6
    
    \bibitem{Amati:1987uf}
    D.~Amati, M.~Ciafaloni and G.~Veneziano,
    \textit{Classical and Quantum Gravity Effects from Planckian Energy Superstring Collisions,}
    Int. J. Mod. Phys. A \textbf{3} (1988), 1615-1661
    doi:10.1142/S0217751X88000710
    
    \bibitem{Kabat:1992tb}
    D.~N.~Kabat and M.~Ortiz,
    \textit{Eikonal quantum gravity and Planckian scattering,}
    Nucl. Phys. B \textbf{388} (1992), 570-592
    doi:10.1016/0550-3213(92)90627-N
    [arXiv:hep-th/9203082 [hep-th]].
    
    \bibitem{Hooft:2015jea}
    G.~'t Hooft,
    \textit{Diagonalizing the Black Hole Information Retrieval Process,}
    [arXiv:1509.01695 [gr-qc]].
    
    \bibitem{Hooft:2016itl}
    G.~'t Hooft,
    \textit{Black hole unitarity and antipodal entanglement,}
    Found. Phys. \textbf{46} (2016) no.9, 1185-1198
    doi:10.1007/s10701-016-0014-y
    [arXiv:1601.03447 [gr-qc]].
    
    \bibitem{Betzios:2016yaq}
    P.~Betzios, N.~Gaddam and O.~Papadoulaki,
    \textit{The Black Hole S-Matrix from Quantum Mechanics,}
    JHEP \textbf{11} (2016), 131
    doi:10.1007/JHEP11(2016)131
    [arXiv:1607.07885 [hep-th]].

    \bibitem{Betzios:2020wcv}
    P.~Betzios, N.~Gaddam and O.~Papadoulaki,
    \textit{Black holes, quantum chaos, and the Riemann hypothesis,}
    [arXiv:2004.09523 [hep-th]].
    
    \bibitem{Gaddam:2020mwe}
    N.~Gaddam and N.~Groenenboom,
    \textit{Soft graviton exchange and the information paradox,}
    [arXiv:2012.02355 [hep-th]].
         
    \bibitem{Regge:1957td}
    T.~Regge and J.~A.~Wheeler,
    \textit{Stability of a Schwarzschild singularity,}
    Phys. Rev. \textbf{108} (1957), 1063-1069
    doi:10.1103/PhysRev.108.1063
     
    \bibitem{Martel:2005ir}
    K.~Martel and E.~Poisson,
    \textit{Gravitational perturbations of the Schwarzschild spacetime: A Practical covariant and gauge-invariant formalism,}
    Phys. Rev. D \textbf{71} (2005), 104003
    doi:10.1103/PhysRevD.71.104003
    [arXiv:gr-qc/0502028 [gr-qc]].
     
    \bibitem{Verlinde:1991iu}
    H.~L.~Verlinde and E.~P.~Verlinde,
    \textit{Scattering at Planckian energies,}
    Nucl. Phys. B \textbf{371} (1992), 246-268
    doi:10.1016/0550-3213(92)90236-5
    [arXiv:hep-th/9110017 [hep-th]].
    
    \bibitem{Zerilli:1970se}
    F.~J.~Zerilli,
    \textit{Effective potential for even parity Regge-Wheeler gravitational perturbation equations,}
    Phys. Rev. Lett. \textbf{24} (1970), 737-738
    doi:10.1103/PhysRevLett.24.737
    
    \bibitem{Nagar:2005ea}
    A.~Nagar and L.~Rezzolla,
    \textit{Gauge-invariant non-spherical metric perturbations of Schwarzschild black-hole spacetimes,}
    Class. Quant. Grav. \textbf{22} (2005), R167
    doi:10.1088/0264-9381/23/12/C01
    [arXiv:gr-qc/0502064 [gr-qc]].
    
    \bibitem{BGP2020}
    P.~Betzios, N.~Gaddam and O.~Papadoulaki,
    \textit{Black hole S-matrix for a scalar field,}
    \textit{to appear.}

    \bibitem{Eichten:1971kd}
    E.~Eichten and R.~Jackiw,
    \textit{Failure of the eikonal approximation for the vertex function in a boson field theory,}
    Phys. Rev. D \textbf{4} (1971), 439-443
    doi:10.1103/PhysRevD.4.439
    
    \bibitem{Tiktopoulos:1971hi}
    G.~Tiktopoulos and S.~B.~Treiman,
    \textit{Relativistic eikonal approximation,}
    Phys. Rev. D \textbf{3} (1971), 1037-1040
    doi:10.1103/PhysRevD.3.1037
    
    \bibitem{tHooft:1989tkg}
    G.~'t Hooft,
    \textit{Black hole Quantization and a connection to String Theory,}
    1989 Lectures, Banff NATO ASI, Part 1, Physics, Geometry and Topology, Series B: Physics, Vol. 238, edited by H.C. Lee, Plenum Press, New York 1990, pp. 105-128.
    
    \bibitem{Kiem:1995iy}
    Y.~Kiem, H.~L.~Verlinde and E.~P.~Verlinde,
    \textit{Black hole horizons and complementarity,}
    Phys. Rev. D \textbf{52} (1995), 7053-7065
    doi:10.1103/PhysRevD.52.7053
    [arXiv:hep-th/9502074 [hep-th]].

    \bibitem{deHaroOlle:1997hx}
    S.~de Haro Olle,
    \textit{Noncommutative black hole algebra and string theory from gravity,}
    Class. Quant. Grav. \textbf{15} (1998), 519-535
    doi:10.1088/0264-9381/15/3/006
    [arXiv:gr-qc/9707042 [gr-qc]].

    \bibitem{deHaro:1998tj}
    S.~de Haro,
    \textit{Planckian scattering and black holes,}
    JHEP \textbf{10} (1998), 023
    doi:10.1088/1126-6708/1998/10/023
    [arXiv:gr-qc/9806028 [gr-qc]].

\end{thebibliography}
\end{document}